\documentstyle[12pt,epsf,twoside,fleqn,espcrc1]{article}
\newcommand{\beq}{\begin{equation}}
\newcommand{\eeq}{\end{equation}}
\newcommand{\beqn}{\begin{eqnarray}}
\newcommand{\eeqn}{\end{eqnarray}}

\newcommand{\AmS}{{\protect\the\textfont2
  A\kern-.1667em\lower.5ex\hbox{M}\kern-.125emS}}
\def\la{\lambda}
\title{Multicritical Matrix Models and the Chiral Phase Transition
\vskip-3cm\hfill\small CPT-98/P.3665\vskip3cm
}

\author{G. Akemann\thanks{Talk presented at the International Symposium 
{\it QCD at Finite Baryon Density}, April 27-30 1998 in Bielefeld, Germany.
To appear in the proceedings Nucl. Phys. A, 
eds. F. Karsch and M.-P. Lombardo.}\\[2ex]
Centre de Physique Th\'eorique CNRS,\\
Case 907 Campus de Luminy, F-13288 Marseille Cedex 9, France}

\begin{document}
\maketitle
\begin{abstract}
Universality of multicritical unitary matrix models is shown and a new
scaling behavior is found in the microscopic region of the spectrum, which  
may be relevant for the low energy spectrum of the Dirac operator at the 
chiral phase transition. 
\end{abstract}

\section {Introduction}
In the last five years analytical results from random matrix theory
have been successfully applied in the investigation of 
Dirac operator eigenvalues provided by QCD lattice data (see \cite{Jac} for
a recent review). A crucial point for their applicability is the question 
of matrix model universality when replacing the Euclidean finite-volume
QCD-partition function by its matrix model counter part:
\beq
{\cal Z} ~=~ \int\! dWd W^{\dagger} {\det}^{N_f}\left( 
              \begin{array}{cc}
              0 & W^{\dagger} \\
              W & 0
              \end{array}
              \right)~ \mbox{e}^{-NTrV(WW^{\dagger})} \ \ ,\ 
V(\la^2)=\sum_{k=1}^p \frac{g_{2k}}{2k} \la^{2k} \ \ .
\label{Zmm}
\eeq
Here the chiral unitary ensemble with $W$ $N\!\times\!(N+\nu)$ complex
has been chosen as an example, which 
corresponds to QCD${}_4$ with $N_f$ massless fermions, $\nu$ zeromodes
and gauge group $SU(N_c\!\geq\!3)$ in the fundamental representation.
It is not fixed apriori which matrix model average $\exp(-NV)$ is to be taken 
since it cannot be derived from the effective Yang-Mills action, that 
averages the determinant of the Dirac operator in QCD.
The matrix model calculation, if applicable, should therefore
be independent of the details of the potential $V$ and thus be universal.
This question has been entirely answered for the unitary ensembles 
in the phase where the chiral symmetry is broken \cite{ADMN,DN}. 
The link between the matrix model eigenvalue correlations and the chiral
condensate is given by the Banks-Casher formula 
$\rho(0)\!=\!\!|<\bar{q}q>\!|/\pi$ 
\cite{BC}. However, the matrix model potential $V$ can be also tuned
in such a way that the eigenvalue density at the origin $\rho(0)$ vanishes. 
These multicritical points \cite{Ka} may therefore serve as a class of models
for the chiral phase transition, where the results presented here mainly 
summarize \cite{ADMNII}. In contrast to \cite{JVS,Hal98} the transition
is not driven by an external parameter introduced to model the effect
of temperature $T$ or finite baryon density $\mu$ in the matrix model.

\section{The chiral phase transition at multicriticality}
There are different large-$N$ limits in which the matrix model eq.(\ref{Zmm})
can be investigated. In the {\it macroscopic limit} $N(\la_i - \la_j)\gg 1$,
where the $\la_i$ are the eigenvalues of $W$,
the eigenvalue density becomes a smooth function with finite support $[-a,a]$ 
\beq
\rho(\la)\ =\ \sum_{k=0}^{p-1} C_k(g_i) \la^{2k} \sqrt{a^2-\la^2}\ \ .
\label{rhomm}
\eeq
It is nonuniversal since the coefficients $C_k(g_i)$
depend explicitly on the couplings of the unknown potential $V(\la^2)$
in the measure (see e.g. \cite{AJM}). Consequently the macroscopic 
density cannot lead to any quantitative prediction on the
Dirac operator spectrum\footnote{The connected density-density
correlator (and all higher correlators) {\it is} universal in the 
macroscopic limit \cite{AJM} as it only depends on $V$
through the support $a$. However, the support of the
Dirac operator eigenvalues is a cut-off dependent quantity,
such that macroscopic universality has no physical implications here.}. 
This has to be compared to the macroscopic density of the 
Dirac operator eigenvalues \cite{Stern93}
\beq
\rho_{Dirac}(\la)\ =\ -\frac{1}{\pi}<\bar{q}q>
\ +\ \frac{(N_f^4-4)<\bar{q}q>^2}{32\pi^2N_fF_\pi^4}|\la| \ +\ o(\la) \ \ ,
\label{stsm}
\eeq
which is clearly a non universal, model dependent quantity as well as
it contains the pion decay constant $F_\pi$ as a physical parameter.

Still, the macroscopic matrix model density carries qualitative information.
If it is identified with the Dirac operator eigenvalue density \cite{SVerb}, 
the chiral condensate enters the matrix model 
through the Banks-Casher relation \cite{BC}. In the microscopic limit 
it will be the only physical parameter for all matrix model correlation 
functions which constitutes their predictive power \cite{Jac}.
The fact that the first coefficient of the macroscopic density
eq.(\ref{rhomm}) $C_0(g_i)\sim <\bar{q}q>$ is vanishing or not
distinguishes between the matrix model being in the symmetric or broken phase.
If the coupling constants $g_i$ are adjusted such that the first
$m\!-\!1$ coefficients vanish, $C_0\!=\!...\!=\!C_{m-1}\!=\!0$
($m$-th multicriticality), a class of possible models for the
chiral phase transition is obtained, where the density vanishes as an 
even power $\rho(\la)\sim \la^{2m}$ at the origin. 
A minimal potential describing such a multicritical behavior is given by
\cite{CM}
\beq
V_m'(\lambda)\ =\ k(m)\lambda^{2m+1}
\left.\left(1 - \frac{1}{\lambda^2}\right)^{1/2}\right|_+ \ \ ,\ \ 
k(m) = 2^{2m+1}\frac{(m+1)!(m-1)!}{(2m-1)!} ~,\label{Vcrit}
\eeq
where the index + indicates to take only positive powers when expanding in 
$1/\la^2$ and $a\!=\!1$ has been chosen. 
The corresponding critical density reads
\beq
\rho_m(\la) ~=~ \frac{1}{2\pi} k(m) \la^{2m} \sqrt{1-\la^2} ~.
\label{rhocrit}
\eeq
The parameter replacing $<\bar{q}q>$ is given by the first non
vanishing term $C_m(g_i)\!\sim\!\rho^{(2m)}(0)$ in eq.(\ref{rhomm}).
The question of universality can
then be addressed by perturbing the minimal potentials (\ref{Vcrit}) by higher 
order terms while maintaining the same critical behavior.
The phase transitions corresponding to these multicritical
points are all of third order as their free energy $F\!=\!1/N^2\ln {\cal Z}$
behaves like \cite{Ka}
\beq
F\ \sim\ (g-g_*)^{2+\frac{1}{m}} \ \ ,\ m\in\mbox{N}\ ,
\eeq
when approaching the critical point.
Investigations using matrix model \cite{Hal98} or renormalization group 
techniques \cite{BR98} (see proceedings) have shown
that the ($T-\mu$)-phase diagram for $<\bar{q}q>$
consists of a first order (small $T$) and a second 
order line (small $\mu$). In the tricritical point where 
the two lines meet the class of transitions provided by the above multicritical
matrix models may become relevant.

\section{A new microscopic scaling limit}
In the {\it microscopic limit} correlations of eigenvalues are considered 
which are in a distance of the order $1/N$. Since the spectrum of the Dirac 
operator close to the origin is of interest a  scaling variable $x\!=\!N\la$ is
defined and kept finite as $N\to\infty$. The rescaled microscopic
eigenvalue density reads
\beq
\rho_S(x)\ \equiv\ \lim_{N\to\infty}\frac{1}{N}\rho(\la=\frac{x}{N})\ \ .
\label{rhomicr}
\eeq
It is a universal function since it depends on the potential $V$ only through
the macroscopic density at zero, $\rho(0)$, and thus it is parameterized by
chiral condensate $<\bar{q}q>$ only. This has been shown for the unitary 
ensembles for an arbitrary polynomial potential at any given 
$N_f$ and $\nu$ for the massless \cite{ADMN} and massive case \cite{DN}. 

In order to extend universality to the multicritical points of 
the unitary ensembles the way to take the microscopic limit has to be modified.
At the phase transition $<\bar{q}q>\!\to\!0$ the 
appropriate scaling behavior for the $m$-th multicritical point 
is found to be \cite{ADMNII}
\beq
N^{\frac{1}{2m+1}}\la \ =\  x \ \ ,\ m\in\mbox{N} \ \ ,\label{scale}
\eeq 
and the microscopic density in terms of the new scaling variable reads
\beq
\rho_S(x)\ \equiv\ \lim_{N\to\infty}N^{-\frac{1}{2m+1}}
\rho(N^{-\frac{1}{2m+1}}x) \ \ . \label{rhomth}
\eeq
The same phenomenon may be expected for the Dirac operator eigenvalues on 
the lattice at the transition. The new universality classes eq.(\ref{rhomth})
will be parameterized by $\rho^{(2m)}(0)$
in analogy to the broken phase \cite{ADMNII}.
Recently the multicritical behavior of a different matrix model 
with random and  deterministic degrees of freedom
has been studied \cite{BH98}. The authors find a modification
of the microscopic scaling at multicriticality as well,
with exponents $ N^{\frac{2k+1}{2k+2}}\la\!=\!x$, $k\!\in$N.
For $k\!=\!1$ such a model has been applied to 
the chiral phase transition \cite{No98} (see proceedings).

\section{Universality at multicritical points}
In this section a brief summary of the analytical and numerical results 
of \cite{ADMNII} is given. 
First the results for the unitary ensemble (QCD${}_3$)
\beq
{\cal Z}\ =\ \int\!dM {\det}^{2N_f}(M) ~e^{-N Tr V(M)} 
    \ \sim\ \int_{-\infty}^{\infty}\! \prod_{i=1}^N \left(d\lambda_i
|\lambda_i|^{2N_f}e^{-NV(\lambda_i)}\right)\left|{\det}_{ij}
\lambda_j^{i-1}\right|^2 ~,
\label{Z}
\eeq
with potential
$V(M)\!=\!\sum \frac{g_{2k}}{2k} M^{2k}$ are given and 
then extended to the chiral unitary ensemble.
In order to determine correlation functions of eigenvalues
a differential equation for the wavefunctions
\beq
\psi_m(\la)\ \equiv\ |\la|^{N_f}
e^{-\frac{N}{2}V(\la)}P_m(\la)\ \ ,\ 
\int_{-\infty}^{\infty}\! d\la~ \psi_m(\la)\psi_n(\la)\ =\ \delta_{mn} ~,
\label{psi}
\eeq
is derived following \cite{KF} and then solved in the microscopic limit 
eq.(\ref{scale}). The $P_n(\la)$ are polynomials
orthonormal to the measure absorbed  in the wavefunctions $\psi_n(\la)$.
The spectral kernel 
\beq
K_N(\lambda,\mu)\ =\ c_N \frac{\psi_N(\mu)\psi_{N-1}(\lambda)
- \psi_N(\lambda)\psi_{N-1}(\mu)}{\mu-\lambda} 
\label{ker}
\eeq
then determines all $n$-point correlation functions from taking its 
determinant. The set of polynomials $P_n(\la)$ obeys the following 
properties
\beqn
\la P_n(\la) &=& c_{n+1}P_{n+1}(\la) ~+~ c_{n}P_{n-1}(\la) ~,\nonumber\\
P_n^{~\prime}(\lambda) &\equiv& A_n(\lambda)P_{n-1}(\lambda) ~-~
                       B_n(\lambda)P_{n}(\lambda) ~,
\label{rec}
\eeqn
where the boundary condition determining the recursion coefficients $c_n$ and
the functions $A_n(\la)$ and $B_n(\la)$
can be found in \cite{KF}. From eqs.(\ref{rec})
the authors derive that the wavefunctions
satisfy the following differential equation for finite $N$
\beq
\psi_n''(\lambda) - F_N(\lambda)\psi_n'(\lambda) + G_N(\lambda)
\psi_n(\lambda) ~=~ 0 ~,
\label{diff}
\eeq
where the $F_N(\la)$ and $G_N(\la)$ are given functions of $A_N$, $B_N$ and 
the potential $V$ as well as derivatives of them \cite{KF}. Taking the 
scaling limit in the broken phase together with the relation 
$\rho(\la)\!=\!\lim_{N\to\infty}
\frac{A_N(\la)}{N\pi}\sqrt{1-(\la/a)^2}$ (\cite{KF}) leads to a
universal differential equation for $\psi_N(x)$ of Bessel type 
first derived in \cite{ADMN}.
When going to the $m$-th multicritical point in the scaling limit
eq.(\ref{scale}) the resulting differential equation will no longer be 
solvable analytically, the $F_N(\la)$ and $G_N(\la)$ being rational functions 
\cite{ADMNII}. 

In the simplest example the $m\!=\!1$-critical potential 
$V(\la)\!=\!-4\la^2 +\frac{1}{4}g\la^4$ 
at $g\!=\!16$ leads to the following coefficients in terms of the 
scaling variable $x\!=\!N^{1/3}\la$:
\beqn
N^{-\frac{1}{3}}F_N(x) &=& \frac{2gx}{u_+ +gx^2} \label {FG} \\
N^{-\frac{2}{3}}G_N(x) &=& \frac{u_+u_-}{4}+((-)^NN_f -\frac{1}{2})v + 
\frac{u_+v+(-)^N2gN_f}{u_+ + gx^2} +\frac{g^2x^4}{4}+ 
\frac{(-)^NN_f -N_f^2}{x^2} \ .\nonumber
\eeqn
The constants $u_{\pm}\!=\! g (2f^2(0) \pm f'(0))$ and $v\!=\!2gf(0)$
have to be determined numerically by solving an auxiliary Painlev\'e II 
equation for the recursion coefficients \cite{CM} at $z\!=\!0$
\beq
0\ =\ gf(z)^3 - zf(z) - \frac{g}{8}f''(z) +\frac{N_f}{2}~. \label{PII}
\eeq 
Still, the issue of universality of the two differential 
equations and thus of $\psi_N(x)$
can be addressed analytically.  Perturbing the critical quartic potential
by a sextic term $g_6\la^6/6$ and maintaining $\rho(\la)\sim \la^2$
all expressions eqs.(\ref{FG}) and (\ref{PII}) remain valid when replacing
$g\to g_*\!=\!g+g_6/2$. Thus $g_*\!=\!\rho''(0)/\pi$ plays the role
of a universal parameter in the solution $\psi_N(x)$ and via the kernel
eq.(\ref{ker}) in all microscopic correlation functions.
Below the numerical solution for the universal microscopic density
eq.(\ref{rhomth}) for $m\!=\!1$ is given. The $x^2$-growth is due to the 
matching condition $\lim_{x\to\infty}\rho_S''(x)\!=\!\rho''(0)$ (dotted line).
\newpage
\begin{figure}[htb]
\begin{minipage}[t]{80mm}
\centerline{\epsfxsize=8cm\epsfbox{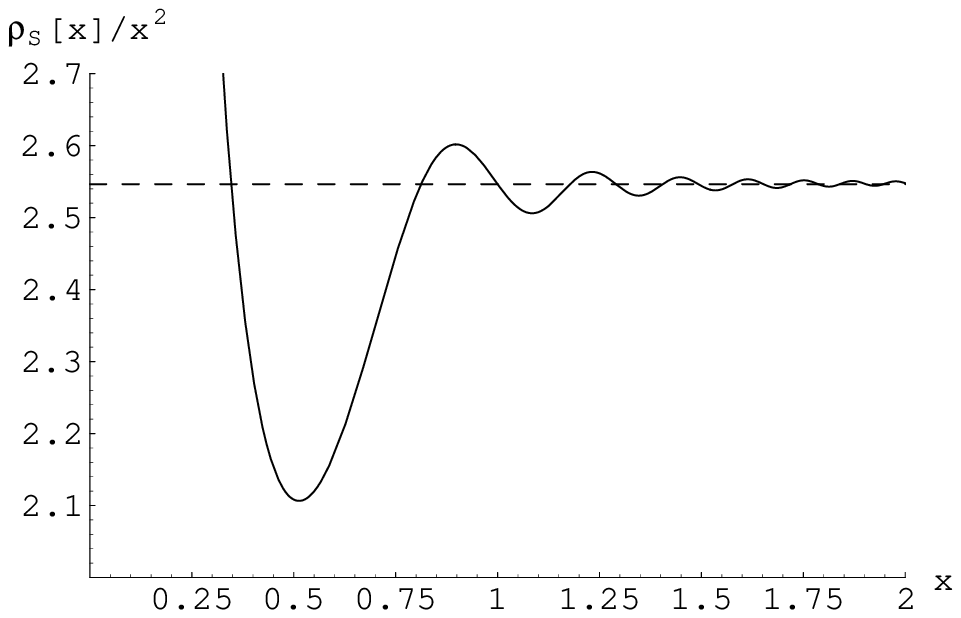}}
\end{minipage}
\hspace{\fill}
\begin{minipage}[t]{75mm}
\centerline{\epsfxsize=8cm
\epsfbox{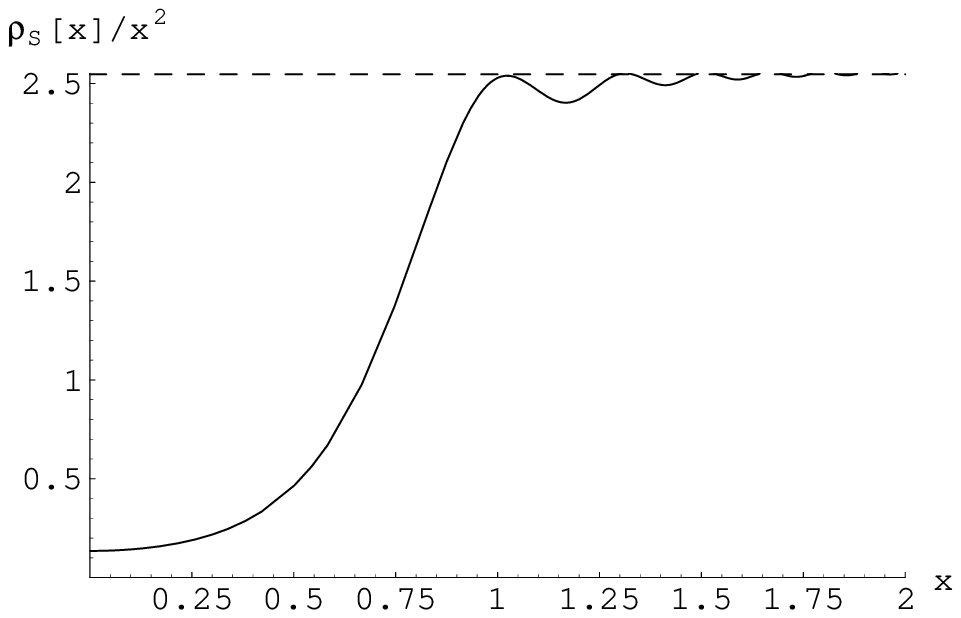}}
\end{minipage}
\end{figure}
Figure 1. The $m\!=\!1$ multicritical density at $N_f\!=\!0$ (left)
and $N_f\!=\!1$ (right). 

\indent

For higher $m$ the functions $F_N(x)$ and $G_N(x)$ become much more involved.
However, when making an approximation to the exact differential equation
(\ref{diff}) in the scaling limit (\ref{scale}) it is possible to obtain
an equation which is analytically solvable for any $m$. Imposing 
\beq
1 \ \ll \ k(m)~x^{2m}  \label{meso}
\eeq
on $x$ for the critical potentials (\ref{Vcrit}) leads to the 
approximate equation
\beq
\psi_N''(x) - \frac{2m}{x}\psi_N'(x) + 
\left(\frac{\pi}{2(2m!)}\rho^{(2m)}(0)x^{4m} 
     + \frac{(-1)^NN_f (2m+1)-N_f^2}{x^2}\right) 
\psi_N(x) \ =\ 0 ,\label{diffmes}
\eeq
which is again of Bessel type. 
Here the constant $k(m)\!=\!2\pi\rho^{(2m)}(0)/(2m)!$ has been replaced. 
The solution reads\footnote{To fix a unique 
solution regularity and normalizability have been imposed as in the broken
phase, although eq.(\ref{diffmes}) is no longer valid at $x\!=\!0$.}
\beq
\psi_N(x)\ \sim\ \sqrt{X}\,
J_{\frac{N_f}{2m+1}-\frac{(-)^N}{2}}(X) \ \ ,\ 
X=\frac{\pi \rho^{(2m)}(0)}{(2m+1)!} ~x^{2m+1} \ \ .
\label{psimeso}
\eeq
The appearance of the parameter $\rho^{(2m)}(0)$ in the 
approximate solutions makes it highly suggestive
to conjecture new universality classes for all multicritical points
with $m\!\geq\!2$ . 

The above results can be easily translated to the multicritical 
chiral unitary ensembles.
The wavefunctions $\tilde{\psi}_n(\la)$ corresponding to the partition
function eq.(\ref{Zmm}) can be simply related to those
of the unitary ensemble above
\beq
\tilde{\psi}_m(\la^2; ~N_f+\nu) \ =\
\psi_{2m}(\la; ~N_f+\nu+\frac{1}{2}) ~~,\label{psich}
\eeq
when shifting $N_f$ in eq.(\ref{Z}) by $\nu+1/2$. 
The approximate analytic solution can thus be 
immediately read off from eq.(\ref{psimeso}). The final
expression for the approximate microscopic density of the 
$m$-th multicritical chiral unitary ensemble reads
\beq
\rho_S(x) =
{\frac{\pi {{\rho }^{(2m)}}(0){x^{2m}}}{2\left( 2m \right) !}}
\left(
{X}\left( 
{{{J_{\beta+{\frac{1}{2}}}}(X)}^2} + 
{{{J_{{\beta-\frac{1}{2}}}}(X)}^2} \right)
-{\frac{2N_f }{2m+1}}
{J_{\beta+{\frac{1}{2}}}}(X)
{J_{\beta-{\frac{1}{2}}}}(X)
\right) \ , \label{finalrhoch}
\eeq
with $\beta\!\equiv\!\frac{2(N_f+\nu)+1}{2(2m+1)}$. It matches 
to the exact noncritical density for $m\!=\!0$ \cite{ADMN}.

\begin{figure}[htb]
\begin{minipage}[t]{80mm}
\centerline{\epsfxsize=8cm\epsfbox{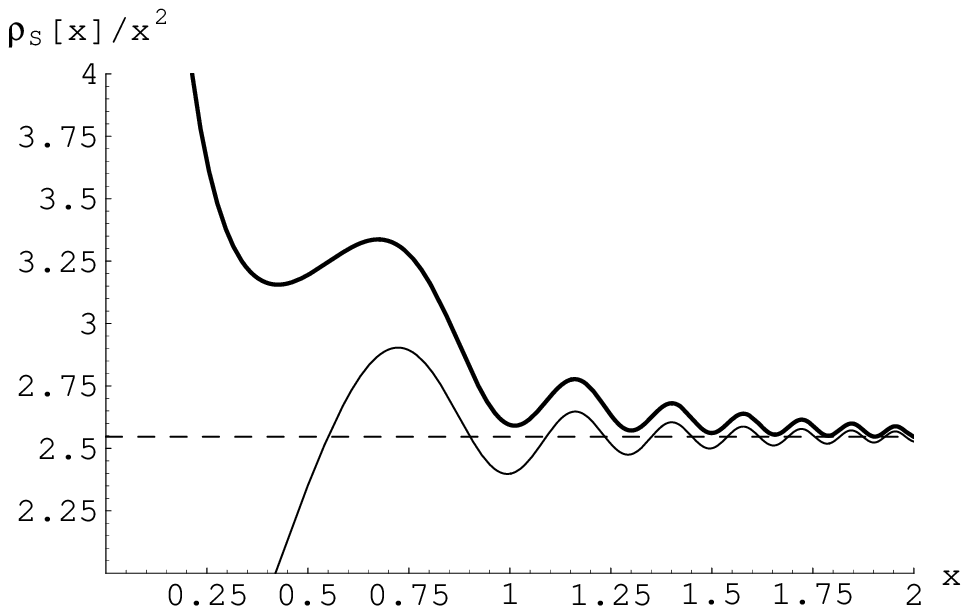}}
\end{minipage}
\end{figure}
Figure 2. A comparison between 
the numerical solution for the $m\!=\!1$ critical microscopic 
density for $N_f\!+\!\nu\!=\!0$ (upper line) and the analytical 
approximation (lower line) which breaks down at small $x$ 
(see eq.(\ref{meso})).

\indent

{\bf Acknowledgments}:
The results of this work were obtained in collaboration with 
P.H. Damgaard, U. Magnea and S.M. Nishigaki. The author is 
supported by the EC grant no. {\small ERBFMBICT960997} and 
wishes to thank the organizers for the very interesting workshop.

\end{document}